\RequirePackage{silence}
\documentclass[fleqn,usenatbib,useAMS]{mnras} 
\usepackage{graphicx}
\usepackage{amsmath}
\DeclareMathOperator{\sign}{sign}
\usepackage{amsfonts}
\usepackage{float}
\usepackage{bm}
\usepackage[perpage,symbol*]{footmisc}
\usepackage{ae,aecompl}
\usepackage{array}
\usepackage{soul}
\usepackage{mathtools}
\usepackage{multirow}

\newcommand{\appropto}{\mathrel{\vcenter{
		\offinterlineskip\halign{\hfil$##$\cr 
			\propto\cr\noalign{\kern2pt}\sim\cr\noalign{\kern-2pt}}}}}

\title{Effect of the Solar dark matter wake on planets} 

\author[Indranil Banik \& Pavel Kroupa]{Indranil Banik$^{1}$\thanks{Email:
\href{mailto:ibanik@astro.uni-bonn.de}{ibanik@astro.uni-bonn.de} (Indranil Banik)\newline $~~~~~~~~~~~~~~~~~~~~$ \href{mailto:pkroupa@uni-bonn.de}{pkroupa@uni-bonn.de} (Pavel Kroupa)} and Pavel Kroupa$^{1,2}$\\
$^{1}$Helmholtz-Institut f\"ur Strahlen und Kernphysik (HISKP), University of Bonn, Nussallee 14$-$16, D-53115 Bonn, Germany \\
$^{2}$Charles University, Faculty of Mathematics and Physics, Astronomical Institute, V Hole\v{s}ovi\v{c}k\'ach 2, CZ-18000 Praha 8, Czech Republic}

\pubyear{2019}
\begin{document}
\label{firstpage}
\pagerange{\pageref{firstpage}--\pageref{lastpage}}

\maketitle

\begin{abstract} 

The Galaxy is conventionally thought to be surrounded by a massive dark matter (DM) halo. As the Sun goes through this halo, it excites a DM wake behind it. This local asymmetry in the DM distribution would gravitationally affect the motions of Solar System planets, potentially allowing the DM wake to be detected or ruled out. Hernandez (2019) recently calculated that the DM-induced perturbation to Saturn's position is 252 metres net of the effect on the Sun. No such anomaly is seen in Saturn's motion despite very accurate tracking of the Cassini spacecraft, which orbited Saturn for >13 years.

Here, we revisit the calculation of how much Saturn would deviate from Keplerian motion if we fix its position and velocity at some particular time. The DM wake induces a nearly resonant perturbation whose amplitude grows almost linearly with time. We show that the Hernandez (2019) result applies only for an observing duration comparable to the ${\approx 250}$ million year period of the Sun's orbit around the Galaxy. Over a 100 year period, the perturbation to Saturn's orbit amounts to <1 cm, which is quite consistent with existing observations. Even smaller perturbations are expected for the terrestrial planets.


\end{abstract}


\section{Introduction}
\label{Introduction}

Over the last century, one of the big `elephants in the room' for astronomers is the fact that very large dynamical discrepancies often exist between the observed rotation curves of galaxies and the predictions of Newtonian gravity applied to their luminous matter distributions \citep[e.g.][]{Babcock_1939, Rubin_Ford_1970, Rogstad_1972}. These acceleration discrepancies are usually attributed to halos of dark matter (DM) surrounding each galaxy \citep{Ostriker_Peebles_1973}. However, the discrepancies follow some remarkable regularities \citep{Famaey_McGaugh_2012} that can be summarised as a unique relation between the acceleration inferred from the rotation curve and that expected from the baryonic distribution \citep{McGaugh_Lelli_2016}. Such a radial acceleration relation was predicted several decades earlier using Milgromian dynamics \citep[MOND,][]{Milgrom_1983}. In this model, the dynamical effects usually attributed to DM are instead provided by an acceleration dependence of the gravity law.

It is important to test both MOND and DM in regimes different to those which gave rise to the paradigms in the first place. For MOND, the dynamics of wide binary stars is a promising `experimentum crucis' in the near future \citep[e.g.][]{Banik_2018_Centauri}. If the observed dynamical discrepancies are due to DM, then massive objects moving through the Galactic DM halo should experience dynamical friction \citep{Chandrasekhar_1943}. This is also true for the Sun, which would be expected to create a trailing DM wake \citep{HERNANDEZ_2019}. Because the DM would be overdense behind the Sun, its distribution would not be spherically symmetric, contrary to the assumption of several previous works \citep[e.g.][]{Pitjeva_2013}. This asymmetry might allow for much stronger constraints on the DM density. Following this idea, \citet{HERNANDEZ_2019} found that the Sun's DM wake would cause the position of Saturn to deviate by 252 metres. Such an effect is not seen in Cassini radio tracking data despite it being accurate to 32 metres \citep{Viswanathan_2017}. This led \citet{HERNANDEZ_2019} to rule out the DM hypothesis.

In this work, we conduct more detailed calculations of how Saturn would deviate from Keplerian expectations if the DM wake is present (Section \ref{Methods}). Our results are shown in Figure \ref{Saturn_displacement} for a century of observations, assuming Saturn's initial position and velocity are known exactly. We derive a perturbation amplitude many orders of magnitude smaller than suggested by \citet{HERNANDEZ_2019}. In Section \ref{Discussion}, we explain why this is. Our conclusions are given in Section \ref{Conclusions}.

\section{Methods and results}
\label{Methods}

\begin{figure}
	\centering
	\includegraphics[width = 8.5cm] {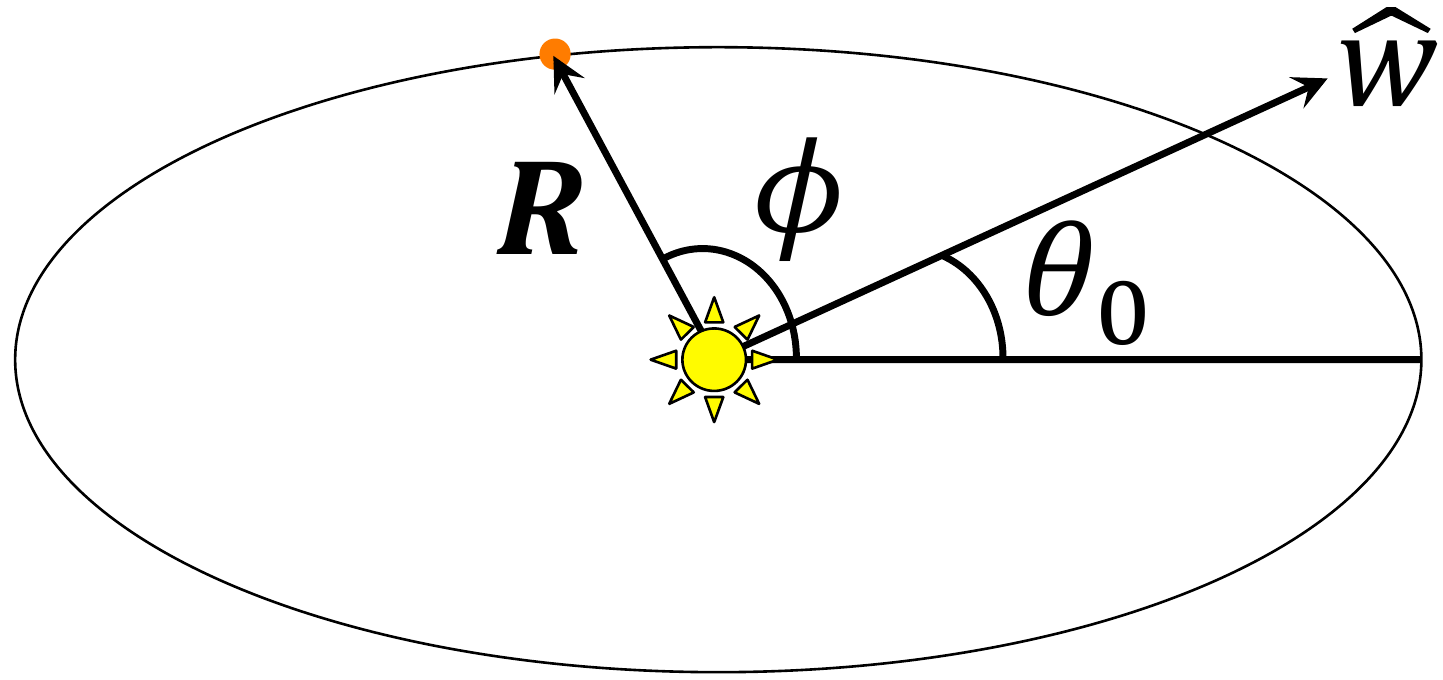}
	\caption{The Solar System geometry assumed in our work. Saturn orbits the Sun on a nearly circular orbit with instantaneous heliocentric position $\bm{R}$, which defines an orbital phase angle $\phi$ relative to some initial direction. We define this to be our $x$-axis and take it to be the direction within Saturn's orbital plane most closely aligned with $\widehat{\bm w}$, the velocity of the Sun with respect to the local DM. A non-zero velocity would cause a density enhancement towards $-\widehat{\bm w}$, the so-called `DM wake'. Its density is axisymmetric with respect to $\widehat{\bm w}$, making the DM-induced tidal acceleration of Saturn very nearly parallel to this direction (see text).}
	\label{Solar_System_diagram}
\end{figure}

In this contribution, we use a co-ordinate system in which the $xy$ plane corresponds to the orbital plane of Saturn, whose orbital pole is along $z$ and instantaneous position is $\bm{R}$. Assuming near-circular motion, we consider the orbit of Saturn in the epicyclic approximation. If its motion were purely Keplerian, its position and velocity would imply that its guiding centre radius (semi-major axis) is $R_0$. At any time $t$, its orbital radius $R \equiv \left| \bm{R} \right|$ is slightly larger by an amount $r$, whose evolution is governed by
\begin{eqnarray}
	\ddot{r} \, + \, \Omega^2 r ~=~ 0 \, ,
\end{eqnarray}
where $\Omega^2 = \frac{GM_\odot}{{R_0}^3}$ is the Keplerian orbital frequency of Saturn and $\dot{q} \equiv \frac{dq}{dt}$ for any quantity $q$.

In Figure \ref{Solar_System_diagram}, we show the relation between Saturn's orbital plane and the DM wake, which we take to lie in the direction $\widehat{\bm w} \equiv \left(\cos \theta_0, 0, \sin \theta_0 \right) \propto \bm v_\odot$. In our notation, $\theta_0$ is the minimum angle between any vector within the orbital plane of Saturn and the velocity of the Sun, $\bm v_\odot$, with respect to the local DM. We assume this DM has no ordered motion in the Galactocentric frame, implying that $\bm v_\odot$ can be measured in this frame.

The Sun moves through the Galactic DM halo with a circular velocity of $v_{c, \odot}$ in addition to some non-circular velocity $\left(U_\odot, V_\odot, W_\odot \right)$. This is defined in the usual Galactic Cartesian frame in which $x$ points towards the Galactic Centre, $z$ towards the North Galactic Pole and $y$ in the direction necessary to make the co-ordinate system right-handed. Fortunately, $y$ points along the local direction of the large scale ordered rotation of the Galactic disk. Thus, the Sun has a Galactocentric velocity
\begin{eqnarray}
	\bm v_\odot ~\equiv~ \begin{bmatrix}
	U_\odot \\
	V_\odot + v_{c, \odot} \\
	W_\odot
	\end{bmatrix} \, .
	\label{Solar_velocity}
\end{eqnarray}
In the rest of this contribution, we use our previously mentioned co-ordinate system aligned with the orbital plane of Saturn. As the direction $\widehat{\bm R} \equiv \bm{R}/R$ towards Saturn changes over the course of its orbit, the angle $\theta$ between $\widehat{\bm w}$ and $\widehat{\bm R}$ is
\begin{eqnarray}
	\cos \theta ~\equiv~ \widehat{\bm w} \cdot \widehat{\bm R} ~=~ \cos \theta_0 \cos \phi \, ,
	\label{cos_theta}
\end{eqnarray}
where $\phi = \Omega t$ is the orbital phase angle of Saturn such that $\bm{R} \equiv R \left( \cos \phi, \sin \phi, 0 \right)$. This is valid for a nearly circular orbit on which $\cos \theta$ has a maximum at $t = 0$. In this case, the component of $\bm R$ parallel to $\widehat{\bm w}$ is
\begin{eqnarray}
	w ~\equiv~ \widehat{\bm w} \cdot \bm R ~=~ R \cos \theta ~=~ R \cos \theta_0 \cos \phi \, .
\end{eqnarray}

The gravity due to the DM wake is approximately parallel to $\bm v_\odot$ \citep{Mulder_1983}. This is due to the large-scale asymmetry caused by the motion of the Sun. Some force is also expected in the direction orthogonal to $\bm v_\odot$, but due to axisymmetry such a force must vanish very close to the Sun. This is not true for a force parallel to $\bm v_\odot$, which we therefore take to be the dominant effect of the DM wake.

After subtracting the gravity exerted by the DM wake on the Sun, the residual acceleration of a test particle close to it can be approximated as \citep[equation A\rm{IV}.13 in][]{Mulder_1983}
\begin{eqnarray}
	\label{Wake_force_general}
	\bm{F}_w &=& -\left(F_1 + F_0 \frac{\cos \theta}{\left| \cos \theta \right| } \right) \widehat{\bm{w}} \, , \, \text{where} \\
	\frac{F_1}{F_{typ}} &=& 0.21 \ln \left(\frac{R}{2r_{min}} \right) \, , \\
	\frac{F_0}{F_{typ}} &=& 0.44 \, , \, \text{with} \\
	F_{typ} &=& \frac{4 \mathrm{\pi} G^2 \rho_{_{DM}} M_\odot}{\sigma^2} \quad \text{and} \\
	\sqrt{\frac{GM_\odot}{r_{min}}} &=& v_\odot \, ,
\end{eqnarray}
where $G$ is Newton's gravitational constant, $\rho_{_{DM}}$ is the unperturbed local density of DM, $\sigma$ is its one-dimensional velocity dispersion, $v_\odot \equiv \left| \bm v_\odot \right|$ and $r_{min}$ is the heliocentric distance at which $v_\odot$ is equal to the circular velocity about the Sun \citep[equation 11 in][]{Mulder_1983}. Note that these results only hold if the unperturbed DM has a Maxwell-Boltzmann velocity distribution with $\sigma = v_\odot/\sqrt{2}$ \citep{HERNANDEZ_2019}. In this case, the most likely DM velocity is $v_\odot \approx 250$ km/s, very close to the values reported for Milky Way analogues in cosmological hydrodynamical simulations \citep[][figure 3]{Bozorgnia_2017}. The numerical coefficients in Equation \ref{Wake_force_general} need to be adjusted for a different $v_\odot/\sigma$ \citep[appendix \rm{IV} in][]{Mulder_1983}.

When estimating the DM wake-induced displacements, we adopt the parameter values listed in Table \ref{Parameters}. At the orbital radius of Saturn, this leads to a typical tidal acceleration of
\begin{eqnarray}
	F_{typ} ~=~ 4.5\times 10^{-21} \, m/s^2 \, .
\end{eqnarray}
This is subject to a ${\approx 10\%}$ uncertainty due to imprecise knowledge of the DM density \citep{Helmi_2018}. Its uncertain velocity distribution also influences our results, with simulations indicating deviations from our assumed Maxwell-Boltzmann form \citep[e.g.][]{Bozorgnia_2017}. There is also a ${\approx 1\%}$ uncertainty in $v_\odot$, corresponding to a ${\approx 2\%}$ uncertainty in $F_{typ}$.

\begin{table}
	\centering
	\begin{tabular}{ccc}
		\hline
		Parameter & Meaning & Value and units \\
		\hline
		R & Orbital radius of Saturn & 9.58 AU \\
		$v_{c, \odot}$ & See Equation \ref{Solar_velocity} & 233.3 km/s \\
		$U_\odot$ & See Equation \ref{Solar_velocity} & 11.1 km/s \\
		$V_\odot$ & See Equation \ref{Solar_velocity} & 12.24 km/s \\
		$W_\odot$ & See Equation \ref{Solar_velocity} & 7.25 km/s \\
		$\rho_{_{DM}}$ & Local dark matter density &  0.018 $M_\odot/pc^3$\\
		\hline
	\end{tabular}
	\caption{Our adopted values of the parameters relevant to this contribution. We obtain $v_{c,\odot}$ from \citet{McGaugh_2018} and the non-circular velocity of the Sun from \citet{Schonrich_2010}. The resulting Solar velocity of 245.9 km/s (Equation \ref{Solar_velocity}) implies that $\theta_0 = 60.6^\circ$ for a planet orbiting within the Ecliptic (Figure \ref{Solar_System_diagram}). The local DM density is constrained by kinematic observations of the Solar neighbourhood \citep{Xia_2016, Helmi_2018}. We assume that the unperturbed DM has a Maxwell-Boltzmann velocity distribution with local one-dimensional velocity dispersion $\sigma = v_\odot/\sqrt{2} = 173.9$ km/s.}
	\label{Parameters}
\end{table}

Over a period of $T = 10$ years, the maximum DM-induced deviation is $\approx F_{typ} \, T^2/2 = 0.22$ mm. Given that the ephemerides of Saturn are accurate to ${\approx 30}$ metres \citep{Viswanathan_2017}, it is unlikely that the motion of Saturn would be noticeably affected by the Solar DM wake. Nonetheless, we determine the expected perturbation more accurately in order to reveal its `resonant' nature \citep{HERNANDEZ_2019}. The precise meaning of this will become clear in the following sections.

\subsection{Motion within the orbital plane}
\label{In_plane_motion}

The force from the DM wake varies only a little for a planet on a near-circular orbit with $R_0 \gg r_{min}$, as is the case for all Solar System planets. Thus, Equation \ref{Wake_force_general} approximately leads to a wake potential within the orbital ($xy$) plane of
\begin{eqnarray}
	\Phi_w &=& \overbrace{F_1 w}^{\Phi_{w,1}} + F_0 \frac{\cos \theta}{\left| \cos \theta \right|} w \\
	&=&  R \cos \theta_0 \left(F_1 \cos \phi + F_0 \left|\cos \phi \right| \right)
	\label{Phi_w}
\end{eqnarray}
On the last line, we restrict attention to the behaviour of $\Phi_w$ within the $xy$ plane.

We now use Chapter 3.3.3 of \citet{Galactic_Dynamics} to consider the evolution of Saturn's orbit in a weakly non-axisymmetric potential. The derivation there requires decomposition of $\Phi_w$ into Fourier modes $\propto \cos m\phi$. The $F_1$ term contributes only to the $m = 1$ mode while the $F_0$ term contributes to all modes with even $m$, including the case $m = 0$. For reasons that we clarify in Section \ref{Higher_harmonics}, we consider only the $m = 0$ and $m = 1$ modes, which we denote $\Phi_{w,0}$ and $\Phi_{w,1}$, respectively. Using equation 3.146 in \citet{Galactic_Dynamics}, we get that
\begin{eqnarray}
    \label{delta_r_governing_equation}
	\ddot{r} + \Omega^2 r &=& -\left( \frac{\partial \Phi_{w,1}}{\partial R} + \overbrace{\frac{2 \Omega \Phi_{w,1}}{R\widetilde{\Omega}}} \right) \cos \widetilde{\Omega}t \nonumber  \\
	&& - \cos \theta_0 F_0 \langle \left| \cos \phi \right|\rangle \, , \text{ where} \\
	\widetilde{\Omega} &\equiv& \Omega - \epsilon \, .
\end{eqnarray}
The term marked with an overbrace arises because the angular momentum oscillates with time, but this effect is only caused by the non-axisymmetric part of $\Phi_w$. To handle the axisymmetric ($m = 0$) part, we use the superposition principle, exploiting the fact that the governing equations are linear for small perturbations. The mean value of $\left| \cos \phi \right|$ is $\langle \left| \cos \phi \right|\rangle = 2/\mathrm{\pi}$. The wake potential oscillates with a very slow frequency $\epsilon \ll \Omega$ due to the Galactic orbit of the Sun. Thus, a very accurate approximation to Equation \ref{delta_r_governing_equation} is
\begin{eqnarray}
	\ddot{r} + \Omega^2 r &=& -\cos \theta_0 \left(3F_1 \cos \widetilde{\Omega}t + \frac{2}{\mathrm{\pi}}F_0 \right) \, .
\end{eqnarray}
We now define scaled force constants
\begin{eqnarray}
	\widetilde{F}_0 &=& -\frac{2 \cos \theta_0}{\mathrm{\pi}} F_0 \, ,\\
	\widetilde{F}_1 &=& -3\cos \theta_0 F_1 \, .
\end{eqnarray}
Thus, we get that
\begin{eqnarray}
	\ddot{r} + \Omega^2 r ~=~ \widetilde{F}_1 \cos \widetilde{\Omega}t + \widetilde{F}_0 \, .
	\label{Governing_equation}
\end{eqnarray}
The particular integral of this can be guessed as
\begin{eqnarray}
	r ~=~ C \cos \widetilde{\Omega}t + \frac{\widetilde{F}_0}{\Omega^2} \, .
	\label{Particular_integral}
\end{eqnarray}
This is a valid solution if the oscillation amplitude is
\begin{eqnarray}
	C ~=~ \frac{\widetilde{F}_1}{\Omega^2 - {\widetilde{\Omega}}^2} \, .
	\label{Forced_harmonic_oscillator}
\end{eqnarray}

Now, suppose we start observing the orbit of Saturn at some time ${t = 0}$. Its position and velocity may have been affected by $\Phi_w$ at earlier times, but we have no way of knowing this due to the lack of prior observations. Thus, all we see is that Saturn behaves as if it has some $r$ and $\dot{r}$ at $t = 0$, which defines a particular (osculating) Keplerian orbit. To detect the DM wake, it is necessary that the behaviour of $r$ deviate from simple harmonic motion satisfying $\ddot{r} + \Omega^2 r = 0$ with $r$ and $\dot{r}$ having their observed values at ${t = 0}$. Thus, we define a further radial perturbation ${\delta r}$ which must satisfy $\delta r = \dot{\delta r} = 0$ when ${t = 0}$. In standard Keplerian motion, ${\delta r = 0} \forall t$.

We have seen that Equation \ref{Particular_integral} solves the governing Equation \ref{Governing_equation}, but it does not satisfy this initial condition. To ensure that it does so, we add appropriate multiples of the complementary functions $\cos \Omega t$ and $\sin \Omega t$.
\begin{eqnarray}
	\delta r ~=~ \frac{\widetilde{F}_0}{\Omega^2} \left(1 - \cos \Omega t\right) \, + \, \frac{\widetilde{F}_1}{\Omega^2 - {\widetilde{\Omega}}^2} \left( \cos \widetilde{\Omega}t - \cos \Omega t \right) \, .
	\label{delta_r_general}
\end{eqnarray}

We now take the limit that ${\epsilon \to 0}$, which implies that ${\widetilde{\Omega} \to \Omega}$. This is valid because $\epsilon$ corresponds to the Galactic orbit of the Sun. Our observations span only a tiny fraction of this ${\approx 250}$ Myr period, making it safe to assume that ${\epsilon t \ll 1}$. In this limit, Equation \ref{delta_r_general} becomes
\begin{eqnarray}
	\delta r ~=~ \frac{\widetilde{F}_0}{\Omega^2} \left( 1- \cos \Omega t \right) \, + \, \frac{\widetilde{F}_1}{2 \Omega} t \sin \Omega t \, .
	\label{delta_r_solution}
\end{eqnarray}

Our derivation could alternatively have been done by assuming ${\epsilon = 0}$ from the outset. In this case, we would need to solve an equation of the form $\ddot{x} + \Omega^2 x = \cos \Omega t$. Given the initial conditions $x = \dot{x} = 0$, the solution to this is $x = t \sin \Omega t/\left( 2 \Omega \right)$, which would yield exactly the same result. Thus, the actual value of $\epsilon$ is completely irrelevant to our analysis as long as $\epsilon$ is sufficiently small that ${\epsilon T \ll 1}$ over the period $T$ covered by accurate observations.

To simplify the notation, we define oscillation amplitudes
\begin{eqnarray}
	x_i ~\equiv~ \frac{\widetilde{F}_i}{\Omega^2} \, , ~i=0,1 \, .
\end{eqnarray}
Combining our previous results, it is easy to see that
\begin{eqnarray}
x_0 &=& -\frac{2 \cos \theta_0 F_0 }{\mathrm{\pi}\Omega^2} = -0.014 \, \text{mm,} \\
x_1 &=& -\frac{3\cos \theta_0 F_1}{\Omega^2} = -0.18 \, \text{mm}.
\end{eqnarray}
In terms of these $x_i$, Equation \ref{delta_r_solution} can be written as
\begin{eqnarray}
	\delta r ~=~ x_0 \left(1 - \cos \phi \right) \, + \, \frac{x_1}{2} \phi \sin \phi \, .
	\label{delta_r_final}
\end{eqnarray}
The $x_i$ are more than just convenient shorthand $-$ they capture the very essence of our whole derivation. If the observing duration is comparable to the orbital period $2 \mathrm{\pi}/\Omega$, we expect to see a perturbation of order ${F_i/\Omega^2}$. Thus, the $x_i$ tell us roughly how large the deviation from Keplerian motion would be after a significant fraction of the planetary orbital period. This also follows from assuming $\delta r \approx F_1 t^2/2$ over a duration $t = 1/\Omega$, a typical orbital timescale. Over longer periods, $\delta r \propto t$ due to the effects of orbital mechanics (Equation \ref{delta_r_final}).

\subsubsection{Tangential motion}

Similarly to ${\delta r}$, we define a perturbation ${\delta p}$ parallel to the circular velocity of Saturn. ${\delta p}$ is governed by equation 3.145 of \citet{Galactic_Dynamics}.
\begin{eqnarray}
	\dot{\delta p} ~=~ -2 \Omega \delta r - \frac{\widetilde{F}_1}{\Omega} \left(\cos \Omega t - 1 \right) \, .
	\label{delta_p_dot}
\end{eqnarray}
As with our solution for $\delta r$, the arbitrary additive constant is chosen to ensure that $\dot{\delta p} = 0$ at $t = 0$. Thus, we get that
\begin{eqnarray}
	\delta p &=& 2x_0 \left(\sin \phi - \phi \right) + x_1 \left( \left(\phi \cos \phi - \sin \phi \right) - \left(\sin \phi - \phi\right) \right) \nonumber \\
	&=& 2x_0 \left(\sin \phi - \phi \right) + x_1 \left(\phi \cos \phi - 2 \sin \phi + \phi \right) \, .
	\label{delta_p_final}
\end{eqnarray}

\subsection{Vertical motion}

Applying Equation \ref{Wake_force_general} in the $z$-direction orthogonal to the orbital plane of Saturn, we get that
\begin{eqnarray}
	F_z ~=~ -\sin \theta_0 \left( F_1 + F_0 \sign \left( \cos \phi \right)\right)
	\label{F_z}
\end{eqnarray}

The oscillatory term involving $F_0$ has a mean value of zero but clearly contributes to the ${m = 1}$ mode, which leads to a resonant perturbation (Section \ref{In_plane_motion}). Its amplitude can be found by Fourier transforming $\sign \left( \cos \phi \right)$, which yields a first non-zero term of $\left( 4 \cos \phi \right)/\mathrm{\pi}$. Neglecting ${m > 1}$ terms as before and noting that the vertical epicyclic frequency is equal to the circular orbit frequency $\Omega$, we get that
\begin{eqnarray}
	\ddot{z} + \Omega^2 z &=& F_{z,0} + F_{z, 1} \cos \Omega t \, , \, \text{where} \\
	F_{z,0} &=& - \sin \theta_0 F_1  = \frac{\tan \theta_0 \widetilde{F}_1}{3} \, ~\text{and} \\
	F_{z,1} &=& - \frac{4 \sin \theta_0 F_0}{\mathrm{\pi}} = 2 \tan \theta_0 \widetilde{F}_0 \, .
\end{eqnarray}
This is directly analogous to the resonant limit of Equation \ref{Governing_equation}, whose solution we have just found (Equation \ref{delta_r_solution}). Therefore, the vertical displacement is
\begin{eqnarray}
	\delta z &=& \frac{F_{z,0}}{\Omega^2} \left( 1- \cos \Omega t \right) \, + \, \frac{F_{z,1}}{2 \Omega} t \sin \Omega t \\
             &=& \tan \theta_0 \left(\frac{x_1}{3} \left(1 - \cos \phi \right) \, + \, x_0 \phi \sin \phi \right) \, .
	\label{delta_z_solution}
\end{eqnarray}

\subsection{The total displacement}

\begin{figure}
	\centering
	\includegraphics[width = 8.5cm] {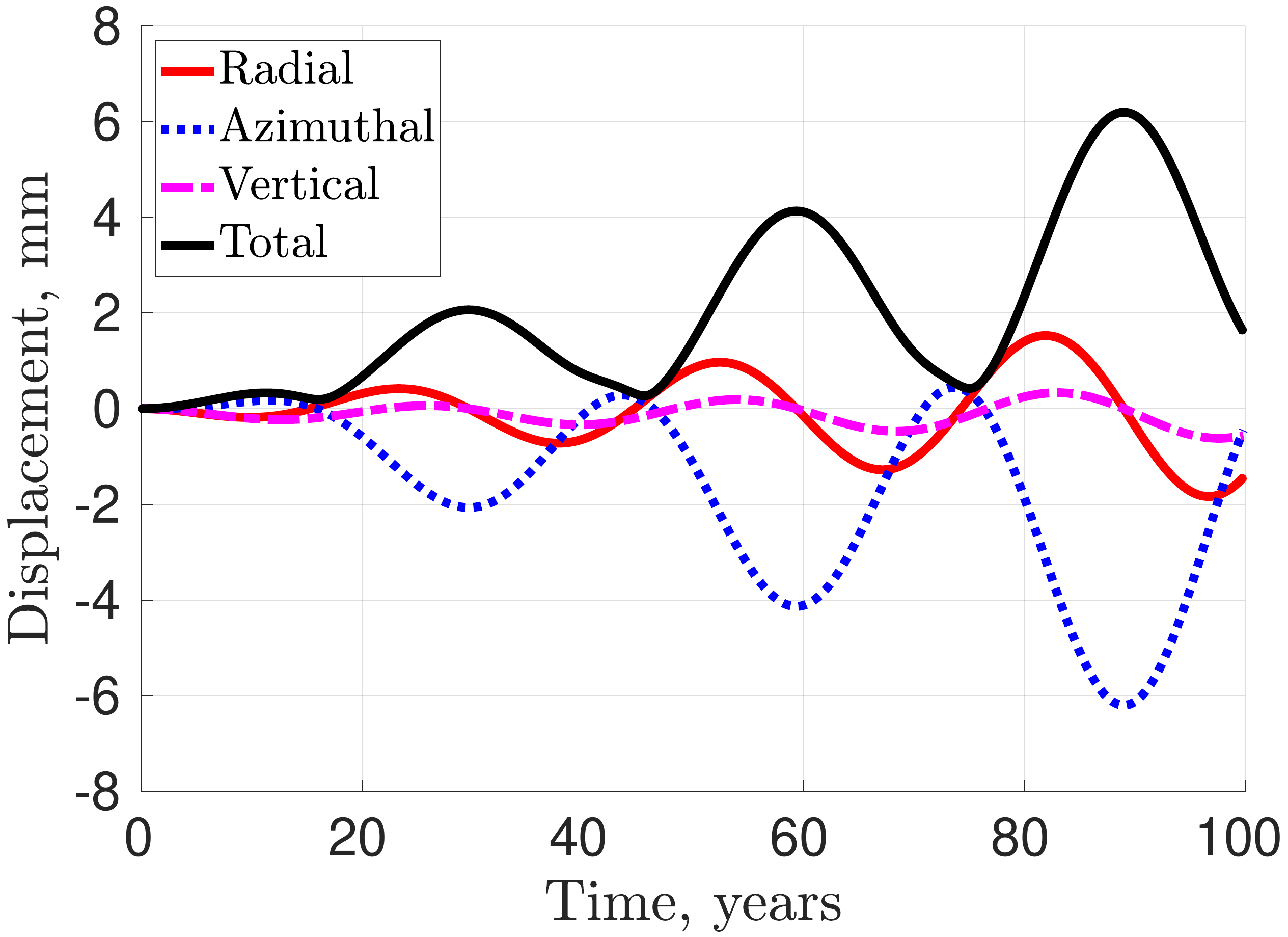}
	\caption{The deviation of Saturn from Keplerian motion due to the Sun's DM wake, assuming a known initial position and velocity. We show the deviation in the radial (solid red), azimuthal (dotted blue) and vertical (dot-dashed pink) directions, with the total shown in black. Observations are assumed to start when the Sun-Saturn line is maximally aligned with the DM wake (${\phi = 0}$ in Figure \ref{Solar_System_diagram}). The perturbations oscillate with an amplitude that grows almost linearly (Section \ref{Methods}). Even so, the total deviation is <1 cm after a century of observations.}
	\label{Saturn_displacement}
\end{figure}

In Figure \ref{Saturn_displacement}, we show the expected deviation of Saturn from Keplerian motion over a period of 100 years. Presently, our observations span approximately half its orbital period of 30 years thanks to the Cassini mission \citep{Matson_1992}. Given an observing accuracy of ${\approx 30}$ metres \citep{Viswanathan_2017}, it is clear that the DM wake cannot be detected using current Solar System observations. Indeed, it would be extremely challenging to detect even if we had similarly accurate observations over a full century.

Observations over such a long period would cover much more than a single planetary orbit, allowing us to approximate that ${\Omega t \gg 1}$. Making this approximation in Equation \ref{delta_r_solution}, we see that the expected displacement ${\propto t/\Omega}$, which is smaller for a less distant planet like Mars. This justifies our focus on Saturn since it is the most distant planet to which we have sent an orbiting spacecraft.

\section{Discussion}
\label{Discussion}

\subsection{Higher harmonics of the wake potential}
\label{Higher_harmonics}

In Section \ref{Methods}, we showed that the ${m = 1}$ mode in $\Phi_w$ leads to a resonant perturbation in the limit that $\epsilon \to 0$. The ${m = 0}$ mode leads to a constant radial gravity which causes Saturn to have a smaller time-averaged $R$ than would otherwise be the case (Equation \ref{delta_r_final}). For a fixed angular momentum, this increases Saturn's average angular velocity, causing ${\delta p}$ to grow linearly with time.

Higher frequency modes $\left( m > 1 \right)$ are also present in Equation \ref{Phi_w}. We now investigate the response to these modes in more detail. For this purpose, Equation \ref{Forced_harmonic_oscillator} is particularly useful since it has a much simpler interpretation in the non-resonant case.

\subsubsection{Radial motion}
\label{Higher_radial_modes}

Equation \ref{Phi_w} leads to ${m > 1}$ modes only because it contains a term of the form $\left|\cos \phi \right|$. We begin by giving the Fourier representation of $\Phi_w$ for modes $m \geq 2$, thus neglecting modes we have already considered in Section \ref{Methods}.
\begin{eqnarray}
	\Phi_w ~=~ -\frac{4 R F_0 \cos \theta_0}{\mathrm{\pi}} \sum_{j = 1}^{\infty} \frac{\left( -1 \right)^j \cos \left(m \phi \right)}{m^2 - 1} \, , \,m \equiv 2j \, .
\end{eqnarray}
Note that only even $m$ modes are non-zero. Neglecting the $m = 0$ mode, Equation \ref{delta_r_governing_equation} now becomes
\begin{eqnarray}
	\ddot{\delta r} + \Omega^2 r ~=~ \frac{12 F_0 \cos \theta_0}{\Omega^2 \mathrm{\pi}} \sum_{j = 1}^{\infty} \frac{\left( -1 \right)^j \cos m\phi}{m^2 - 1} \, , \,m \equiv 2j \, .
\end{eqnarray}
Applying the standard solution for a forced harmonic oscillator (Equation \ref{Forced_harmonic_oscillator}) and adding appropriate multiples of the complementary function to satisfy the boundary condition $\delta {r} = \dot{\delta {r}} = 0$ at ${\phi = 0}$, the solution is
\begin{eqnarray}
	\delta r ~=~ 6 x_0 \sum_{j = 1}^{\infty} \frac{\left( -1 \right)^j \left(\cos m \phi - \cos \phi \right)}{ \left( m^2 - 1 \right)^2} \, , \,m \equiv 2j \, .
	\label{delta_r_higher}
\end{eqnarray}
Comparing this with the dominant $x_1$ term in Equation \ref{delta_r_final}, we can now estimate the fractional error in that equation due to neglecting $m > 1$ modes.
\begin{eqnarray}
	\text{Fractional error in } \delta r ~\approx~ \frac{12 \, x_0}{ \left( m^2 - 1 \right)^2 x_1} \approx 0.10 \left( m = 2 \right) \, .\nonumber
\end{eqnarray}

\subsubsection{Tangential motion}
\label{Higher_tangential_modes}

The generalisation of Equation \ref{delta_p_dot} to ${m > 1}$ is
\begin{eqnarray}
	\dot{\delta p} ~=~ -2 \Omega r - \frac{\Phi_w}{R\Omega} \left( \cos m\phi - 1\right) \, .
\end{eqnarray}
Substituting in our solution for ${\delta r}$ due to modes with ${m > 1}$ (Equation \ref{delta_r_higher}) and applying our usual boundary conditions, this becomes
\begin{eqnarray}
	&&\dot{\delta p} ~=~ -\sum_{j = 1}^{\infty} \left[\frac{2 \Omega x_0 \left( -1 \right)^j}{\left(m^2 - 1\right)^2} \right] \\
	&& \left[6 \left(\cos m \phi - \cos \phi \right) \right. + \left(m^2 - 1\right) \left(\cos m\phi - 1\right)\left.\right] \, , \,m \equiv 2j \, .\nonumber
\end{eqnarray}
Requiring $\delta p = 0$ fixes the solution to
\begin{eqnarray}
	&&\delta p ~=~ -\sum_{j = 1}^{\infty} \frac{2 \, x_0 \left( -1 \right)^j}{\left(m^2 - 1\right)^2} \nonumber \\
	&&\left[ \left(m^2 - 5 \right) \frac{\sin m\phi}{m} - 6 \sin \phi - \left(m^2 - 1 \right) \phi \right] \, .
	\label{delta_p_higher}
\end{eqnarray}
Similarly to Section \ref{Higher_radial_modes}, we can estimate the fractional effect of ${m > 1}$ modes on ${\delta p}$ by comparing Equations \ref{delta_p_final} and \ref{delta_p_higher}. In the limit that $\phi \gg 1$, we get that
\begin{eqnarray}
	\text{Fractional error in } \delta p ~\approx~ \frac{2 x_0}{ \left( m^2 - 1 \right) x_1} \approx 0.05 \, \left( m = 2 \right) \, .\nonumber
\end{eqnarray}

\subsubsection{Vertical motion}
\label{Higher_vertical_modes}

Equation \ref{F_z} induces ${m > 1}$ modes due to the $\sign \left( \cos \phi \right)$ term, which yields only odd $m$ modes. By finding the strengths of these Fourier modes and neglecting the previously considered case ${m = 1}$, we get that
\begin{eqnarray}
	\ddot{\delta z} + \Omega^2 z ~=~ -\frac{4 F_0 \sin \theta_0}{\mathrm{\pi}} \sum_{j = 1}^\infty \frac{\left(-1 \right)^{j + 1}}{2j - 1} \cos \overbrace{\left ( 2j - 1 \right)}^{m} \phi \, .
\end{eqnarray}
Following our derivation in Section \ref{Higher_tangential_modes}, the solution is
\begin{eqnarray}
	\delta z ~=~ \sum_{j = 1}^\infty \frac{2 \, x_0 \tan \theta_0 \left(-1 \right)^{j + 1}}{m \left(m^2 - 1 \right)} \left(\cos m \phi - \cos \phi \right) \, .
\end{eqnarray}
Comparing this with Equation \ref{delta_z_solution} shows that it has an error from $m > 1$ modes of
\begin{eqnarray}
	\text{Fractional error in } \delta z ~\approx~ \frac{6 \, x_0}{ m \left( m^2 - 1 \right) x_1} \approx 0.02 \, \left( m = 3 \right) \, .\nonumber
\end{eqnarray}

\subsubsection{Combined effect}
\label{Combined_higher_modes}

Figure \ref{Saturn_displacement} indicates that ${\delta p}$ is responsible for the largest part of the total DM-induced displacement. Our results in Section \ref{Higher_tangential_modes} indicate that ${m > 1}$ modes only affect ${\delta p}$ by ${\approx 5\%}$. Similarly small effects can be expected for the sub-dominant contributions from ${\delta r}$ (Section \ref{Higher_radial_modes}) and ${\delta z}$ (Section \ref{Higher_vertical_modes}). Thus, the total DM-induced displacement of Saturn is only affected a few percent by $m > 1$ harmonics in $\Phi_w$. This is much smaller than uncertainties from other parameters like the ${\approx 10 \%}$ error in the local DM density \citep{Helmi_2018}. Therefore, it is currently not very important to consider these higher harmonics.

\subsection{Comparison with \citet{HERNANDEZ_2019}}
\label{Comparison_with_Hernandez}

Our results in Figure \ref{Saturn_displacement} are much smaller than those obtained by \citet{HERNANDEZ_2019}. This is because we properly take the resonant limit of Equation \ref{delta_r_general} while \citet{HERNANDEZ_2019} found the maximum amplitude of the induced perturbations. This maximum is reached after order the Galactic orbital period of the Sun. For a much shorter observing duration, the maximum possible amplitude of $\delta r$ is not relevant. The fact that this is only a few hundred metres \citep{HERNANDEZ_2019} shows just how little the DM wake would really influence the Solar System.

The difference between our approach and that of \citet{HERNANDEZ_2019} is perhaps most evident in his equation 6, which states that the perturbation is order $\widetilde{F}_1/\left(\epsilon \Omega\right)$ in our notation. In reality, our observations span a duration $T \approx 1/\Omega$, so the maximum deviation that we could see due to the $\widetilde{F}_1$ term is $\widetilde{F}_1 T^2/2 \approx \widetilde{F}_1/\left(2\Omega^2\right)$. \citet{HERNANDEZ_2019} obtained a much larger result because he implicitly assumes that our observations are long enough to cover all relevant oscillation periods, including the Galactic orbit of the Sun. This overestimates the observing duration by $\approx 7$ orders of magnitude, explaining why our calculated perturbation amplitudes are smaller than his by roughly this factor.

Nevertheless, we emphasize the great importance of using the Solar System as a laboratory for testing fundamental physics. Radio tracking data from Cassini has placed strong constraints on the MOND interpolating function, though some choices are still consistent with observations \citep{Hees_2014, Hees_2016}. Nearby wide binary stellar systems could be even more important as the MOND circular velocity exceeds the Newtonian value by ${\approx 20\%}$ \citep{Banik_2018_Centauri}. Such systems do appear to show a departure from Newtonian expectations \citep{Hernandez_2018}, though it was later shown that the two can be reconciled with a more careful rejection of outliers \citep{Banik_2019_Centauri}.

The Chandrasekhar dynamical friction expected from DM remains a promising way to show its reality, independently of direct and indirect detection experiments. Although the effects are small in the Solar System, galactic-scale tests are much more promising since the amount of DM in a system roughly scales with the cube of its size, more than compensating for the inverse square law of Newtonian gravity. Results from several galactic tests argue against the reality of the putative DM halos \citep{Angus_2011, Kroupa_2015, Oehm_2017}. More accurate observations of the Pisces Overdensity could shed light on this issue if it is indeed a feature in the Galactic stellar halo caused by the orbit of the Large Magellanic Cloud \citep{Belokurov_2019}.

\section{Conclusions}
\label{Conclusions}

The hypothetical DM wake behind the Sun would have effects on planets within the Solar System \citep{HERNANDEZ_2019}. However, that work overestimates the effect due to implicitly assuming that our observations cover a duration comparable to the Galactic orbital period of the Sun. This is most evident in his equation 6, which shows a larger deviation than the product of force and observing duration squared. Over the finite ($\la 30$ year) period covered by accurate space-age observations, the DM wake of the Sun has an imperceptible effect on the motion of Solar System planets. Thus, the DM hypothesis is not in tension with presently available Solar System ephemerides. Galactic-scale tests appear more promising, with the nearby M81 group providing evidence against the expected dynamical friction \citep{Oehm_2017}. In addition, wide binary systems should soon clarify the true cause of the observed dynamical discrepancies in galaxies.

\section*{Acknowledgements}

IB is supported by an Alexander von Humboldt postdoctoral fellowship. The graph was prepared using \textsc{matlab}$^\text{\textregistered}$.

\bibliographystyle{mnras}
\bibliography{DFW_bbl}
\bsp
\label{lastpage}
\end{document}